\documentclass[twocolumn,aps,pre,showpacs,amsmath,amsfonts,amssymb,floatfix]{revtex4}
\usepackage{morefloats}
\usepackage{amsmath}
\usepackage{graphicx}
\usepackage{dcolumn}
\usepackage{bm}

\begin{document}

\title{Assortative and disassortative mixing investigated using the spectra of graphs}

\author{Sarika Jalan $^{1,2}$ \email{sarika@iiti.ac.in} and Alok Yadav${^1}$}
\affiliation{1. Complex Systems Lab, Indian Institute of Technology Indore, M-Block, IET-DAVV Campus, Khandwa Road, Indore 452017, India}

\affiliation{2. Centre for Biosciences and Biomedical Engineering, Indian Institute of Technology Indore, M-Block, IET-DAVV Campus, Khandwa Road, Indore 452017, India}

\begin{abstract}
We investigate the impact of degree-degree correlations on the spectra of networks. Even though density distributions exhibit drastic changes depending on the (dis)assortative mixing and the network architecture, the short range correlations in eigenvalues exhibit universal RMT predictions. The long range correlations turn out to be a measure of randomness in (dis)assortative networks. The analysis further provides insight in to the origin of high degeneracy at the zero eigenvalue displayed by majority of the biological networks.
\end{abstract}

\maketitle

\section{Introduction}
Last two decades have witnessed a rapid advancement in the field of complex 
networks \cite{Rev_Barabasi,Boccaletti2006,Barabasi_2004}. The prime idea governing this framework is to consider a system made of 
interacting units. To categorize and understand real world systems based on interacting units, many models have been proposed, among which Erd\"{o}s-R\'{e}nyi (ER) random \cite{ER}, scale-free (SF) \cite{SF} and small world \cite{SW} are the most popular ones. Further, degree-degree correlations have also been used as one of 
the key properties of networks characterization
\cite{Newman,Boccaletti2006,Rivera2010,Newman2003,Barthelemy_2003,Newman2003c,Assort_Ref,Soc_Bio_Assort,Newman_ref,Croft,Bollen,Bagler,Bassentt2008}
and is known to confer robustness to biological networks \cite{Plos_one_trans}.
The tendency of (un)like degree nodes to stick together is termed as (dis)assortativity.
Various social networks are known to be assortative while few of the biological and 
technological networks have been reported to be disassortative 
\cite{Soc_Bio_Assort,Newman_ref,Croft,Bollen,Bagler,Bassentt2008}. Despite its importance for real networks, (dis)assortativity does not appear in any of the model networks discussed above, and is driven by some other mechanism, for example reshuffling algorithm \cite{Igor}. While spectral behaviour of uncorrelated networks have been quite well understood \cite{degree}, despite real world systems being highly correlated \cite{Newman2003}, such understanding for the correlated networks still needs to be developed.
  
Spectral graph theory is an established branch of mathematics, and eigenvalues of corresponding adjacency matrices are known as finger prints of the underlying graphs \cite{Van_book,Akemann_book,Chung_book,Rowlinson_book}. With recent advancement in the network theory, the spectral graph theory, traditionally used in investigations of random and regular graphs, got extended to studies of graphs motivated by real world systems.
These spectral studies, apart from presenting bounds for extremal eigenvalues highlight their importance by relating them with the various structural as well as dynamical properties of the networks \cite{largest_virus,largest_syn}.
The studies of networks further reveal a key impact of assortativity on the extremal eigenvalues \cite{PVan}, which has been explored in context of disease spreading \cite{Goltsev} and diffusion processes \cite{Caldarelli2012}, thereby exhibiting the importance of spectral studies of networks for a more comprehensive understanding of complex systems.
This paper presents a systematic analysis of impact of
degree-degree correlations on the spectral properties of various networks under the random matrix theory (RMT) framework.
Since its introduction in 1960s, in the context of nuclear spectra, the theory has been
successfully applied to a wide range of complex systems ranging from the quantum chaos to galaxy \cite{rev_RMT1,rev_RMT2}. Recently, 
with a spurt in the activities of network framework, the RMT got its extension in
analysis of spectral properties of various model networks 
\cite{Vattay,SJ_PRE_2007b} as well as those arising from real world systems \cite{prl_amino,SJ_PlosOne}.

\section{Methods and Techniques}
To quantify the degree-degree correlations of a network, we consider the Pearson (degree-degree) correlation coefficient, given as \cite{Newman,Newman2003}
 \begin{equation}
r = \frac{[M^{-1}\sum_{i=1}^{M} j_i k_i] - [ M^{-1}\sum_{i=1}^{M} \frac{1}{2}(j_i + k_i)^2]}
{[M^{-1}\sum_{i=1}^{M} \frac{1}{2}(j_i^2+ k_i^2)] - [ M^{-1}\sum_{i=1}^{M} \frac{1}{2}(j_i + k_i)^2]}, 
\label{assortativity}
 \end{equation}
where $ j_i, k_i $ are the degrees of nodes at both the ends of the $i^{th}$ connection and $M$ represents the total connections in the network.

The random network of size $N$ and average degree $\langle k \rangle$ is constructed 
using the ER model by connecting each pair of nodes with the probability $p = \langle k \rangle/N$ \cite{ER}. These networks have assortativity coefficient ($r$) being close to zero or exactly zero. To generate the networks with various assortativity, we use the reshuffling algorithm \cite{Igor}. In this algorithm, after selecting two pairs of nodes randomly, we sort them according to degree. The highest degree node is then connected to the second highest degree node with the reshuffling probability $p_r$, which governs the (dis)assortative mixing, i.e. we reconnect a high degree node to a (low) high degree one and low degree  node to a (high) low degree one. With the probability $1-p_r$, we rewire them randomly. If new connection resulting from this rewiring 
already exists, it is discarded and the previous steps are performed. The process is 
carried out until a steady value of $r$ is attained. For assortative networks, the $k$ degree nodes to form a complete graph with the value of $r$ being one, the network should have at least $(k+1+2n)$ nodes, where $n$ can be any integer starting from 0. As this condition is not satisfied for all the degrees present in the network, the network takes a value lesser than one. Similarly, the disassortative network can have the value of $r$ less than $-1.0$. 

We make a further note that at high assortativity values, all the 
similar degree nodes being connected among themselves form groups \cite{Igor}. 
As we decrease the assortativity, the connections within the groups of similar degree nodes decrease 
and the connections between different groups of similar degree nodes increase. 
For disassortative networks, connections between different groups of similar degree nodes exist giving rise to a bipartite-like structure \cite{Igor}.

The SF networks of size $N$ and average degree $\langle k \rangle$ are generated using the Barab\'asi-Albert algorithm by starting with a completely connected network seed and adding new nodes one by one which connect with existing nodes using the preferential attachment method \cite{Rev_Barabasi}.

The networks are represented in the form of adjacency matrix by defining $A_{ij}=1$, if $i$ and $j$ nodes are connected otherwise 
$A_{ij}=0$. 
For an undirected and unweighted network with N nodes, the adjacency matrix is $N \times N$ symmetric square matrix entailing all real eigenvalues. We denote the eigenvalues as 
$\lambda_i, i=1,2..N$ and $\lambda_{i}\leq \lambda_{i+1}$ and analyse them under the RMT framework. The random matrix studies consider two properties of a spectra: (1) global properties such as spectral distribution of the eigenvalues
$\rho(\lambda)$, and (2) local properties such
as eigenvalue fluctuations around $\bar\lambda$. In RMT, calculations of spectral fluctuations are done using the unfolded eigenvalues $\bar{\lambda_i}= \bar{N}(\lambda_i)$, where $\bar{N}(\lambda)=\int_{\lambda_{min}}^{\lambda}\rho(\lambda\acute{})\,d{\lambda\acute{}}$ is the average integrated eigenvalue density \cite{Mehta_book}. By using these unfolded
eigenvalues, nearest neighbour spacings are calculated as $s_i = \bar{\lambda}_{i+1} - \bar{\lambda_{i}}$.
For symmetric random matrices with the mean zero and the variance one, the 
nearest neighbour spacing distribution (NNSD) follows GOE statistics given as:
\begin{equation}
P(s)=\frac{\pi}{2}s \exp(-\frac{\pi s^2}{4}),
\label{eq_GOE}
\end{equation}
which shows a level repulsion at small spacing values with an exponential fall for larger spacings
indicating that nearest neighbour eigenvalues are correlated \cite{Mehta_book}. Whereas the spacing distribution of a matrix whose diagonal elements are 
Gaussian distributed random 
numbers and rest of the elements are zero exhibit Poisson statistics $(P(s) = \exp(-s))$ indicating that eigenvalues are uncorrelated \cite{Mehta_book}. 

The intermediate of these two distributions can be characterized using the Brody equation \cite{Brody}:
\begin{equation}
P_{\beta}(s)=A s^\beta\exp\left(-\alpha s^{\beta+1}\right),
\label{eq_brody}
\end{equation}
where $A$ and $\alpha$ are determined by the parameter $\beta$ as $A =(1+\beta)\alpha$ and 
$\alpha=\left[{\Gamma{\left(\frac{\beta+2}{\beta+1} \right) }}\right]^{\beta+1} $.  The value of Brody parameter lies in the range $(0\leq \beta \leq 1)$. The value of $\beta$ being $0$, indicates the Poisson distribution, where as $\beta=1$ corresponds to the GOE distribution. Other
values of $\beta$ indicates that the distribution lies intermediate to these two. 

The NNSD provides a correlation measure of subsequent eigenvalues, whereas the $\Delta_3(L)$
statistic measures how the eigenvalues which are $L$ distance apart are
correlated, and can be estimated using the least-square deviation of the spectral 
staircase function representing average integrated eigenvalue density 
$\bar{N}(\lambda)$ from the best fitted straight line for a finite interval of 
length $L$ of the spectrum given by \cite{rev_RMT2}:
\begin{equation}
\Delta_3(L;x)=\frac{1}{L}\min_{a,b}\int_x^{x+L}[N(\bar{\lambda})-a\bar{\lambda}-b]^2\,d\bar{\lambda}
\label{delta}
\end{equation} 
where $a$ and $b$ are regression coefficients obtained after least square fit. Average over several choices
 of x gives the spectral rigidity, the $\Delta_3(L)$.
For the GOE statistics, the $\Delta_3(L)$ depends on $L$ in the following manner: 
\begin{equation}
\Delta_3 (L)\sim \frac{1}{\pi^2} \ln L
\label{eq_delta}
\end{equation}
For the network spectra considered in this paper, there is no analytical form of $\bar{N}$, and we perform unfolding by numerical 
polynomial fitting using the smooth part of the spectra by discarding eigenvalues towards both the ends as well as degenerate eigenvalues, if any \cite{rev_RMT1,rev_RMT2}. 
This renders the dimension of the unfolded eigenvalues less than the dimension of the network.

\section{Results}
\begin{figure}[b]
\includegraphics[width=\columnwidth, height=7.7cm]{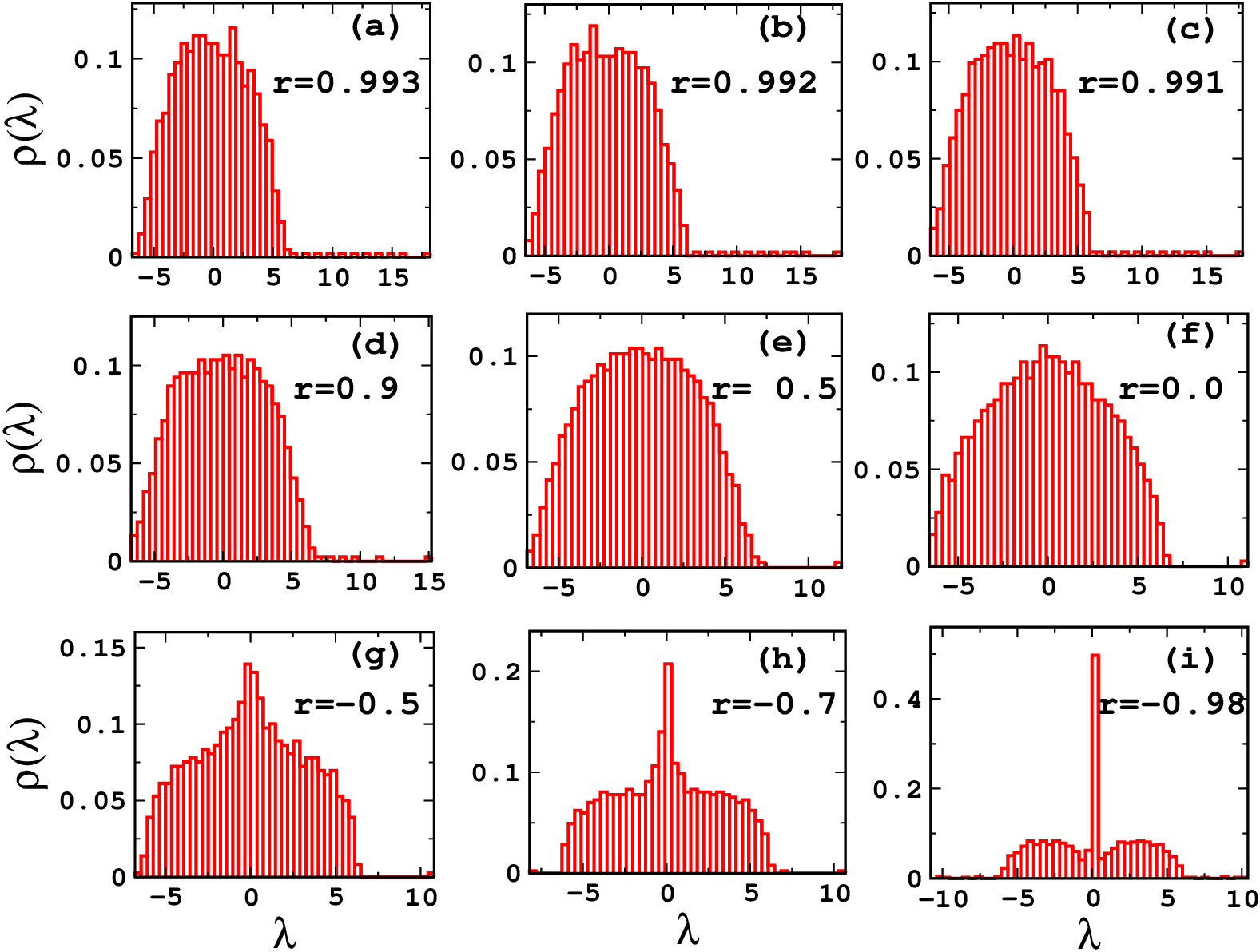}
\caption{(Color Online) Spectral density
for Erd\"os-R\'enyi random networks with different values of assortativity coefficient $r$. All graphs are plotted
 for
 the networks with size $N=1000$ and connection probability $p=0.01$, averaged over twenty different realizations of the networks.}
\label{fig_density_rand}
\end{figure}

The bulk part of the spectra of ER random networks with $r$ value being close to zero, follow the well known semi-circular law \cite{Farkas_PRE_2001,Aguiar} (Fig.~\ref{fig_density_rand}(f)). The extremal eigenvalues deviate from the random matrix predictions and indeed
provide various information about structural and dynamical properties of corresponding systems
\cite{largest_virus,largest_syn,largest_eigval,Goltsev,Newman2003c}. 
In the following, we present results pertaining to the impact of assortativity on the spectral properties of networks. 
It turns out that with an increase in the assortativity, the semi-circular distribution, as observed for the uncorrelated ER random networks, remains unchanged (Fig.~\ref{fig_density_rand}(a)-(e)). The largest eigenvalue exhibits an increasing trend as already discussed in \cite{PVan,Goltsev}. 
As network is rewired entailing disassortativity, spectral distribution ($\rho(\lambda)$) acquires a very different structure than those of the assortative networks. The 
networks start exhibiting a high degeneracy at zero, with overall spectra resembling a double humped structure 
(Fig.~\ref{fig_density_rand}(h)), which becomes more pronounced as the disassortativity becomes higher or value of $r$ becomes more negative (Fig.~\ref{fig_density_rand}(i)). This increase in disassortativity is also accompanied with more number of degenerate eigenvalues at zero. There could be various reasons for this high degeneracy, few of them, appropriate in the present context are:
First, as discussed that disassortativity supports bipartite-like structure \cite{Igor} and a complete bipartite network has all zero eigenvalues except two. Hence bipartite-like behaviour of the disassortative networks presents one of the reasons for the occurrence of
high degeneracy at zero. Second, tree-like structure has been demonstrated to yield degeneracy
at zero eigenvalue \cite{tree} and disassortativity encourages tree-like structure \cite{Igor},
which in turn indicates high degeneracy at zero. 
We remark that for large $N$, the limiting shape of $\rho(\lambda)$ is known
for various cases, which for sufficiently dense matrices, tend to follow the Wigner semi-circular law 
typical for the Gaussian matrix ensembles \cite{Farkas_PRE_2001,Aguiar}, 
whereas an ensemble of sparse random matrices of finite size are known to yield states beyond the semi-circular law in the 
tails of the distribution \cite{Rodgers,Mirlin,Evangelou}. 
For sparse random graphs, i.e matrices with $0$ and $1$ entries having smaller $p$
values, while the density distribution 
$\rho(\lambda)$ of an ensemble exhibit singularities, with
the height of the peaks being the corresponding multiplicities, the bulk is still shown to comply with random matrix predictions of Wigner's semicircular law \cite{Rogers,Dumitriu}.
Moreover, investigations of various model networks mimicking real world properties have revealed that the spectra of these networks
exhibit degeneracy at zero \cite{Mehta_book}, as observed for the sparse random matrices. On that account, despite degeneracy at zero, the bulk of the assortative networks following the semi-circular distribution, is not surprising.
\begin{figure}[t]
\includegraphics[width=\columnwidth, height=7.8cm]{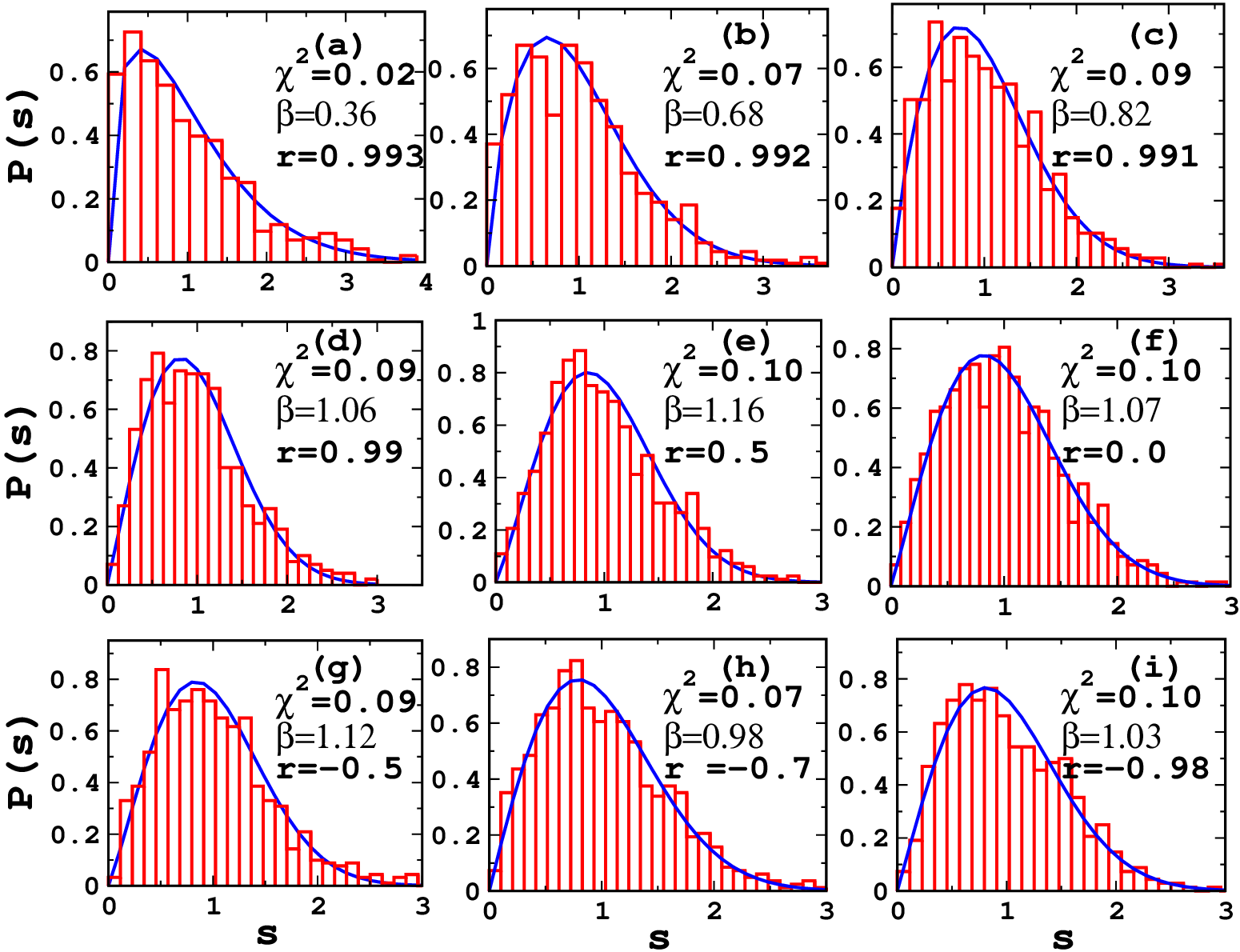}
\caption{(Color Online) The NNSD for Erd\"os-R\'enyi random networks with different values of assortativity coefficient $r$. All graphs are plotted for the networks with size $N=1000$ and connection probability $p=0.01$. Histograms are from the data points and solid line is for fitting with Brody distribution (Eq.~\ref{eq_brody}). }
\label{fig_spacing_rand}
\end{figure}

As the spectral density only provides a global behaviour of eigenvalues, in order to get insight into local fluctuations, we further analyse the short range and long range correlations in eigenvalues. The NNSD follows GOE statistics of RMT (Eq.~\ref{eq_GOE}) for all the values of $r$ except for very high values corresponding to the highly assortative networks (Fig.~\ref{fig_spacing_rand}). What is interesting that the values of $r$ for which $\rho(\lambda)$ exhibits a very similar behaviour, except a change in the value of the largest eigenvalue, the NNSD captures crucial structural changes reflected through the value of the Brody parameter. For the highest achievable value of the assortativity coefficient for the particular network parameter for which results are presented, the value of $\beta$ comes out to be close to $0.3$ (Fig.~\ref{fig_spacing_rand}(a)), and as the assortativity decreases we witness a smooth transition to the GOE statistics with value of the $\beta$ turning one. 
Depending upon the network size, average degree and degree sequences, the highest achievable value of $r$ for that network may be different (as discussed in the Section II), which might lead to a different value of $\beta$. Fig.~\ref{fig_density_rand}(a)-(d) depict that a very small change in the value of $r$ is capable of entailing a profound change in the statistics, in-fact it approaches from the Poisson to the GOE. Since a very small randomness is known to be enough in introducing the short range correlation in eigenvalues \cite{SJ_pre2007a}, for a very small deviation from the highest assortativity entails GOE statistics. Since the assortativity in network supports a the groups having similar degree nodes and as assortativity decreases, these distinct groups of nodes observed for very high values of $r$ gets destroyed leading to a transition from the Poisson to the GOE statistics. As soon as the value of $r$ is decreased, and sufficient random connections among the groups of similar degree nodes are induced, the value of Brody parameter $\beta$ becomes one and no further signature of structural changes on value of $\beta$ is found with a further decrease in the assortativity.
\begin{figure}[b]
\includegraphics[width=\columnwidth, height= 6.0cm]{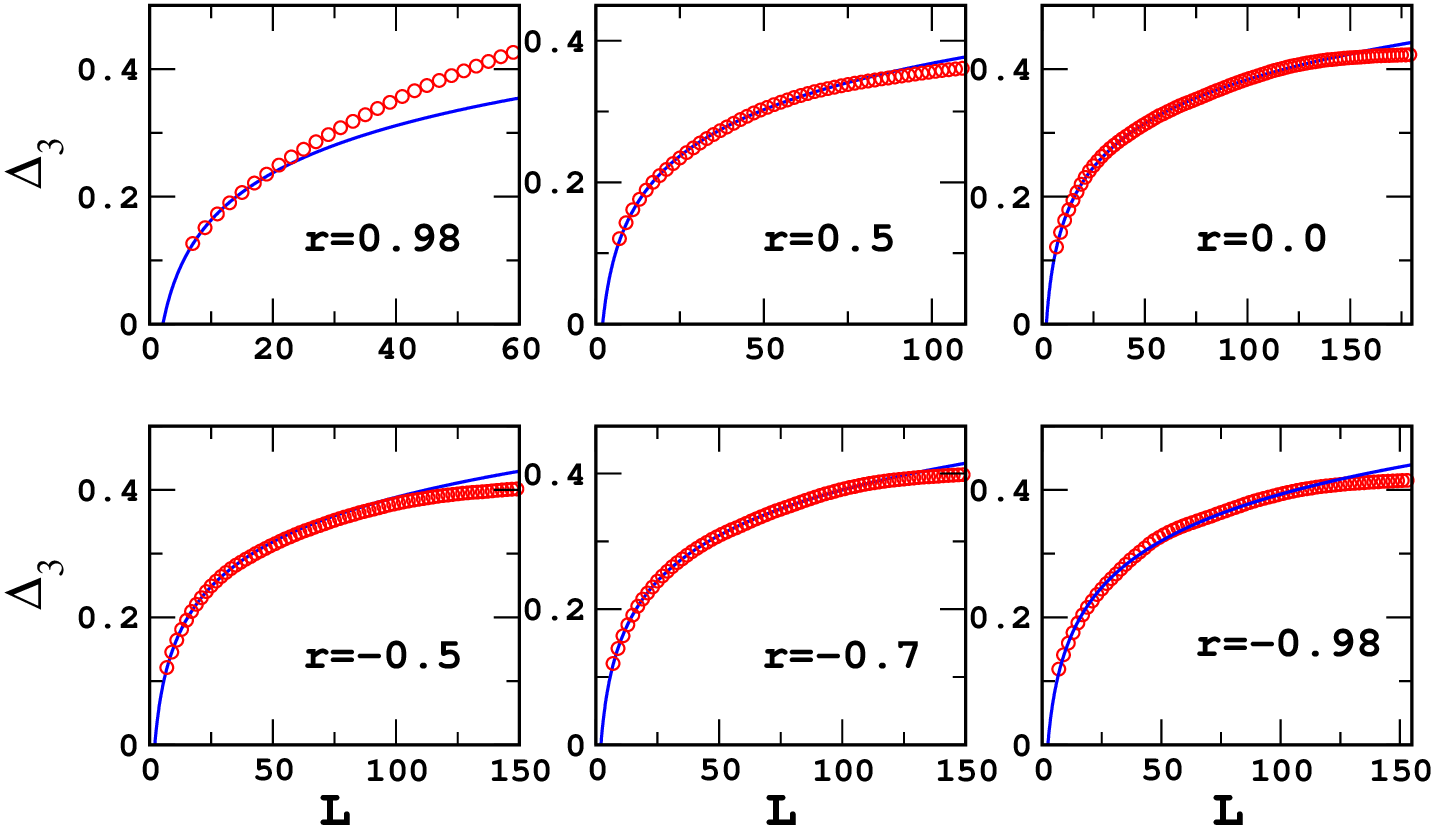} 
\caption{(Color Online) The $\Delta_3(L)$ statistic for Erd\"os-R\'enyi random networks with different values
of assortativity coefficient $r$. All graphs are plotted for the networks with size $N=1000$ and connection probability $p=0.01$. Solid line is the prediction from GOE statistics (Eqs.~\ref{delta} and \ref{eq_delta}) and open circle are calculated from the network.}
\label{fig_delta}
\end{figure}

For disassortative 
networks which are characterized with negative values of $r$, what is remarkable is that despite these networks displaying distinguishable spectral distributions than those of the assortative networks, the NNSD yields the value of the Brody parameter ($\beta = 1$) bringing them into the universality class of GOE. This is not surprising as NNSD 
is analysed by taking the non-degenerate part of the spectra, and high degeneracy at a particular value, for instance at zero, does not account for any effect in the NNSD. As long as the underlying network has some random connections, the NNSD displays the GOE statistics \cite{SJ_pre2007a}. 
We remark that all the networks considered here form a single connected cluster as for disconnected networks, even though each individual sub-network follows GOE statistics, the spectra taken together may lead to a different spacing statistics \cite{Vattay}.  

In order to get a further insight to the structural changes arising due to the changes in $r$ values, we probe for the long range correlations in eigenvalues for those sets which yield the $\beta$ value one. 
We find that for all these values of $r$, the long range correlations, measured using the $\Delta_3(L)$ statistic (Eq.~\ref{delta}) follow the universal GOE statistics as given by Eq.~\ref{eq_delta} for a certain value of $L$ (denoted as $L_0$) and deviates from this universality afterwards (Fig.~\ref{fig_delta}). 
\begin{figure}[b]
\includegraphics[width=\columnwidth, height=7.7cm]{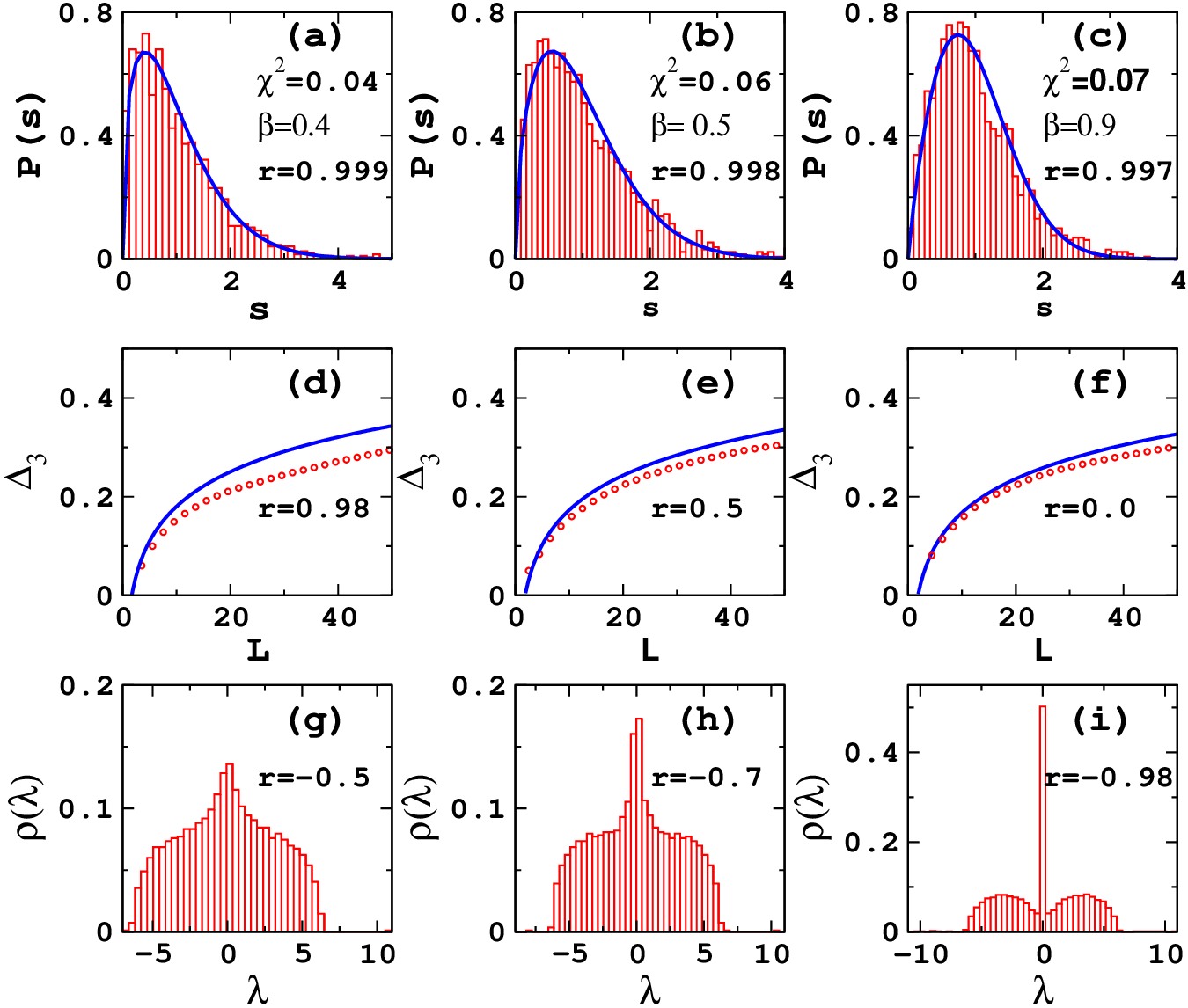}
\caption{(Color Online) (a)-(c) represent the NNSD, (d)-(f) present the $\Delta_3(L)$ statistic and (g)-(i) depict the spectral density distribution of ER random networks.
All graphs are plotted
 for the networks with size $N=2000$, $\langle k \rangle$=10 and for average over twenty different realizations of the network.}
\label{figN2000}
\end{figure}

\begin{figure}[t]
\includegraphics[width=\columnwidth, height=7.7cm]{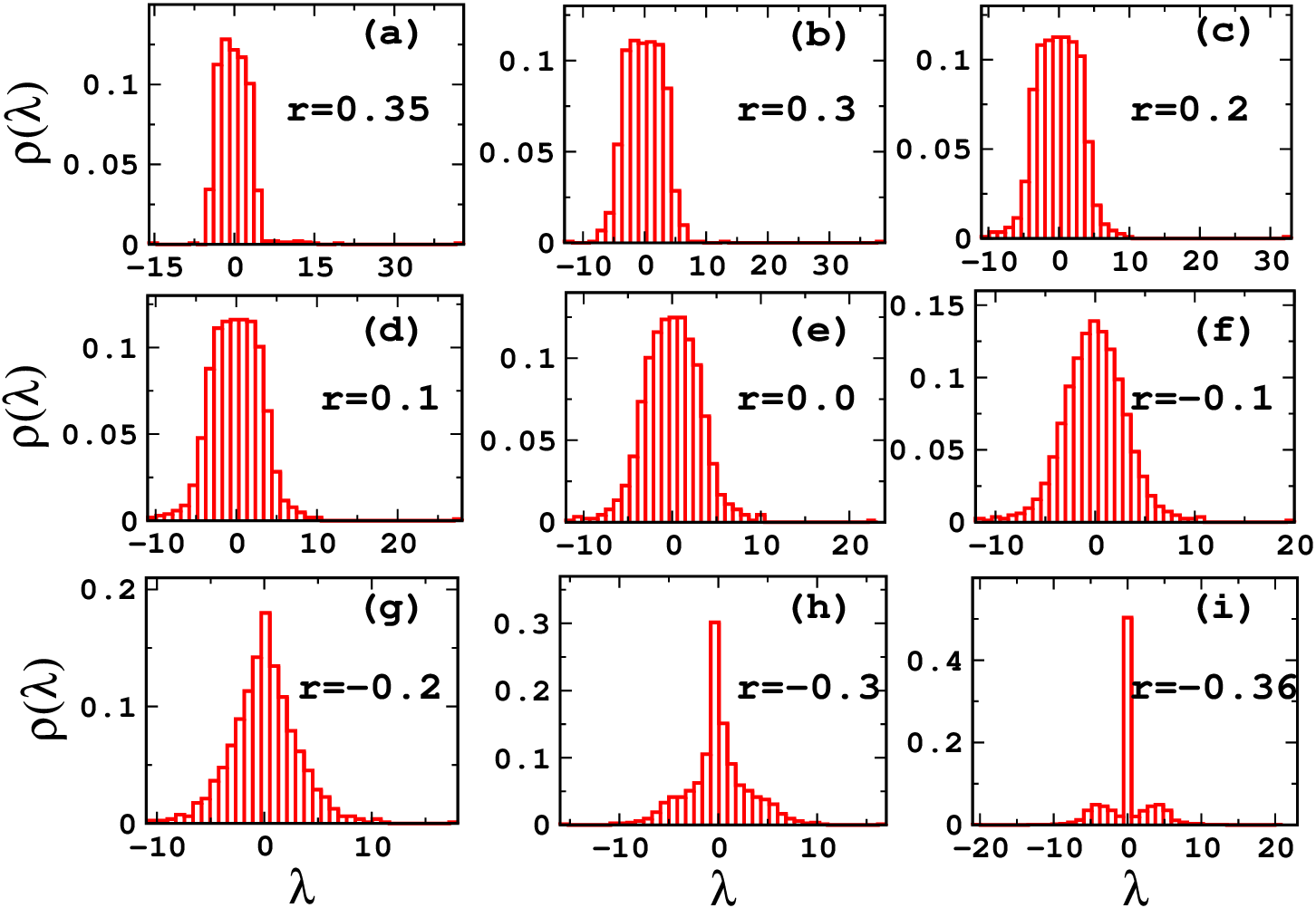}
\caption{(Color Online) Spectral density 
for scale-free networks with different values of assortativity coefficient $r$. All graphs are plotted
 for the networks with size $N=1000$, $\langle k \rangle$=10, for twenty different realizations.}
\label{fig_density_SF}
\end{figure} 

Note that a regular network, for instance 1-d lattice with a periodic boundary condition, follows Poisson distribution. 
As connections are rewired, thereby increasing the randomness in the network, value of the Brody parameter increases with 
an increase in the rewiring probability and becomes one at the onset of the 
small-world transition, demonstrating that nearest neighbour eigenvalues are correlated \cite{SJ_pre2007a}. For such a small change in the network structure there is no visible change
in the density distribution, but the Brody distribution detects even such a small change in
the number of random connections, and hence has been proposed to be used as a measure of
randomness at a fine scale \cite{SJ_pre2007a}. After the Brody parameter attains a value one, the $\Delta_3(L)$ 
statistic has been shown to measure the randomness (in terms of $L_0$ in this paper), in the underlying network \cite{SJ_epl2009}.As rewiring probability
increases further, the value of $L_0$ for which $\Delta_3(L)$ statistic follows RMT predictions increases, demonstrating that the eigenvalues which are $L_0$ distance apart are also correlated. Since $L_0$ provides a measure of randomness in a network \cite{SJ_epl2009}, for the networks under investigation in the present work, it turns out that the highest assortative network is least random, as value of $L_0$ is least for that particular $r$ value (Fig.~\ref{fig_delta}(a)). As assortativity of the network is decreased, the randomness of the network increases reflected in the higher value of $L_0$. This increase in the size of $L_0$ continues up to $r$ being zero, supporting the fact that network reaches to the maximum randomness. The value of $L_0$ then remains steady for a further decrease in the value of assortativity to the minimum possible value of $r$, i.e. to the maximum disassortativity (Fig.~\ref{fig_delta}(c)). As most of the real world networks have been reported to posses certain level of disassortativity \cite{Newman2003}, based on the $\Delta_3(L)$ results we can argue that real world systems attempt to have more randomness, thereby leading to be disassortative. What follows that as value of $r$ increases, by keeping network size and average degree same,  the value of $L_0$ for which $\Delta_3(L)$ statistic
follows RMT predictions increases, indicating an increased amount of randomness in the underlying network. Fig.~\ref{figN2000} demonstrates that the behaviour of various spectral properties remain unchanged as network size increases. Figs.~\ref{figN2000}(a)-(c) indicate that the value of Brody parameter $\beta$ becomes one with a very small decrease in the value of $r$. With a further decrease in the value of $r$, the value of $L_0$ for which the $\Delta_3$(L) statistic follow GOE statistic increases indicating an increase
in the randomness as discussed earlier. With a further decrease in the value of $r$ in the disassortativity regime, there occurs a peak at zero eigenvalue which becomes more pronounced as network becomes more dissociative which is also accompanied with the deviation from
the semi-circular distribution at very low value of $r$.

Further, in order to demonstrate the robustness of the universal RMT predictions against changes in the network architecture, we present results for the SF networks for various values of $r$. For $r$ being close to zero, the density distribution of SF networks exhibit the triangular shape \cite{Aguiar}, which, with an increase in the assortativity, tends to display flattening of the peak. The range of the distribution also shrinks as the assortativity increases (Fig.~\ref{fig_density_SF}(a)-(e)). On the other hand, as we decrease assortativity, i.e. make the network more disassortative, the shape of density distribution starts changing from 
its signature triangular distribution, with the peak at zero eigenvalue being more pronounced (Fig.~\ref{fig_density_SF}(f)-(g)). As we further increase the disassortativity, the 
eigenvalues distribute themselves symmetrically and adopt a double-hump shape for highly disassortative networks, clearly visible in Fig.~\ref{fig_density_SF}(i) which is accompanied by a high peak at the zero eigenvalue similar to that of the ER random networks. It is noteworthy that for highly disassortative networks, the spectral density of ER and SF model networks behave similarly, deviating from their respective signature distributions.
Further, the $\beta$ value exhibits a transition from the Poisson to the
GOE statistics with a decrease in the $r$ value. Despite the overall spectral density being different from that of the ER networks, the NNSD and $\Delta_3(L)$ statistic display similarity in behaviour, which is in line of the argument that the eigenvalues fluctuations are calculated from the smooth homogeneous part of the spectra by not taking degeneracy into account and density is not known to be a real test of GOE statistics \cite{Pandey_1983}. 
 
\begin{table}[t]
\label{table_config_networks}
\begin{center}
    \begin{tabular}{ | c | c | c | c | c | c |}
    \hline
PPI networks   & N  & $r_{0}$  &$N_0(PPI)$& $N_0(r=0)$ & $ N_0(r=r_0)$  \\ \hline 
{\it H.pylori  }      & 709  & -0.243 & 317 & 115   & 152               \\ \hline
{\it C.elegans }      & 2386 & -0.183 & 1354 & 465   & 1124              \\ \hline 
{\it S.cerevisiae }   & 5019 & -0.088 & 976 & 717   & 1149              \\ \hline 
{\it H.sapiens }      & 2138 & -0.084 & 864 & 423   & 643               \\ \hline 
{\it D.melanogaster } & 7321 & -0.083 & 2311 & 1389  & 1975              \\ \hline 
{\it E.coli }         & 2209 & -0.012 & 487 & 487   & 497               \\ \hline 
    \end{tabular}
\end{center}
\caption{Comparison of number of zero eigenvalues of PPI networks of different species and their corresponding configuration models. $r_{0}$ denotes the value of the assortativity coefficient for the PPI networks. $N_0(PPI)$ denotes the number of zero eigenvalues in the spectra of the PPI networks. $N_0(r=0)$ stands for
degeneracy at zero for configuration model with $r=0$, whereas $N_0(r= r_0)$ denotes the same for the configuration models taking
$r$ values equal to the corresponding PPI network.}
\end{table}

We would like to remark here on the impact and reliability of network size considered in the present investigation. 
In RMT, different quantities are calculated by averaging an ensemble of matrices. 
However for real systems, calculations are made as running averages over part of the 
whole spectrum. The random matrix predictions can be applied to real world systems if above 
two are equivalent, a property known as ergodicity. More explicitly, it means that all 
members of the ensemble, except for a set of measure zero, satisfies the above 
equivalence \cite{ergodicity1,ergodicity2}. Due to the ergodicity, one can construct matrix ensembles in different ways: (a) large dimensional random matrices with less number of realizations, or (b) smaller dimensional matrices with large number of realizations. We consider an ensemble of twenty network realizations with a large dimension, which is already shown to be good enough to study various structural properties of networks, such as degree distributions, clustering coefficients etc. \cite{SW}. Moreover, individual entities of each ensemble 
follow RMT predictions for NNSD with a good accuracy, characterized by $\chi^2$ values. As we increase the realizations, accuracy increases (Fig.~\ref{fig_spacing_rand} and Appendix). Consideration of an ensemble consisting of many more number of network realizations would not lead to significant betterment or difference in the following properties of the network spectra: (1) the Brody parameter smoothly turning one with a decrease in the value of $r$ at a very fine scale; (2) a further decrease in the values of $r$ leading
to an increase in the value of $L$ for which spectra follows GOE statistics; and (3) increasing height of the peak at zero eigenvalues with an increase in the disassortativity, owing to the bipartite-like structure of the network. 

Next, in order to investigate if the 
degree-degree correlations in real world system have different spectral behaviour
than those of the model networks discussed above, we consider the
 protein-protein interaction (PPI) networks of six different species. These networks have already been shown to follow universal RMT predictions of GOE statistics \cite{PPI}. We concentrate here on the occurrence of high degeneracy at the zero eigen value. The assortativity coefficient and fraction of degenerate eigenvalues are tabulated in Table I. As all the PPI networks possess negative value of $r$ as well as have a high degeneracy at zero, we expect disassortativity to be one of the factors governing the degeneracy in the real world networks. In order to probe more into the correlation between disassortativity and degeneracy at zero, we compare the corresponding configuration model for all PPI networks presented above (Table I). It is clearly indicated that as soon as the value of $r$ takes a negative value (close to the corresponding PPI network), while keeping all other parameters of the system same, there is an increase in the degeneracy at zero eigenvalue.

\section{Discussion and Conclusions}
The density distribution of the random networks for $r$ value being zero follow the Wigner semi-circular distribution. Even with change in the assortativity ($0 \le r < 1$), 
the bulk part of the spectra keeps displaying semi-circular distribution (Fig.~\ref{fig_density_rand}(a)-(f)), whereas increase in the disassortativity ($-1\le r < 0$) leads to the double hump, which is symmetrically distributed around a peak at zero eigenvalue (Fig.~\ref{fig_density_rand}(i)). The height of the peak increases with the increase in the disassortativity of the network.
 
The NNSD of the networks with the various (dis)assortativity values ($1 < r < -1$), reveal that there is a smooth transition in the $\beta$-value around the very high assortativity regime. For 
very high assortativity values, $\beta$ values lie close to zero, and as network becomes less assortative $\beta$ progresses to one.
It might be due to the reason that the networks with the 
highest assortativity has groups of similar degree nodes which get perturbed as $r$ decreases by making random connections among these different groups. For rest of the assortativity values, the $\beta$ remains fixed at $1$, which corresponds to the universal GOE distribution as $r$ value goes to negative end. 
 
Further, the property of Brody parameter being able to detect changes in network structure at a fine scale and the increase in $L_0$ of $\Delta_3(L)$ statistic after attainment of $\beta$ value one have several implications, one of them that concerns the present work is that the value of $\beta$ distinguishes two networks based on random connections present, while the other is that more assortativity in the network corresponds to less randomness. Decreasing the assortativity 
leads to an increase in randomness which continues up to the value of $r=0$, for which the network is most random ($L_{0}$ value being maximum). 
Then the value of $L$ for which $\Delta_3(L)$ statistic follows GOE prediction starts decreasing and remains steady for a further decrease in the value 
of assortativity up to the minimum possible value of $r$ (i.e. up to the maximum disassortativity case). The SF networks also exhibit the similar 
statistics of eigenvalue fluctuations as for ER random networks, where density distribution for $r=0$ and for lower $ \vert r \vert$ values 
show triangular distribution instead of semi-circular. Both the networks, however, exhibit high degeneracy at zero for the disassortative networks. 
By considering different PPI networks, we further demonstrate the role of disassortativity governing the appearance of degeneracy at zero eigenvalue. 
 
In spectral graph theory, most of the works concentrate on extremal eigenvalues 
\cite{Van_book}; whereas the RMT research focuses on distribution of various spectral properties of random matrices with an extension to the random graphs,
largely ignoring many graphs properties existing in real world systems. The analysis
carried here is a step towards bridging this
gap by considering two most popular tools of random matrix theory, i.e. density
and spacing distributions, to understand the impact of one of the important properties of graphs i.e. assortativity.  
This property has been increasingly realized as a characteristic of a system
\cite{Tenenbaum,Gaytan,realnetwork}. Our analysis is another demonstration 
of the importance of spacing analysis in understanding impact of degree-degree
correlation on the network detected through the spectra
as for very minute changes in $r$, there are no visible changes in
the spectral density, but this leads to a very drastic changes in the eigenvalues
fluctuations demonstrating the impact of $r$ values on randomness in a network.

Furthermore, $\Delta_3(L)$ statistic provides an insight into the reason why social networks tend to be assortative while biological and 
technological networks tend to be disassortative. As randomness, measured in terms of $L_0$ for which the $\Delta_3(L)$ 
statistic follows RMT prediction, increases with a decrease in $r$. A direct implication of this result can be witnessed in case of social 
networks where entities are known to be associated in ordered fashion (people with similar age or educational profile are often more connected) \cite{homophily}, thus providing a probable reason as to why social networks tend to assume an assortative topology. On the other hand, it has been reported that most of the biological and technological networks 
possess a certain level of disassortativity \cite{Newman,Bassentt2008,Newman2003}. Also the biological networks, for instance the PPI networks exhibit varying amounts of randomness in their underlying networks detected through different values of $L_0$ for which the $\Delta_3(L)$ statistic follows GOE statistics \cite{PPI}. This randomness has been attributed to mutations occurring in course of evolution \cite{mutation}. Relating the disassortative nature of the PPI networks and the randomness they possess, with the results obtained from our analysis of the model networks, suggests that biological networks tend to become more disassortative in order to comply with their underlying randomness. 

To conclude, we present a systematic analysis of the spectral 
properties of the networks with varying (dis)assortativity. 
We find that assortativity has a profound impact on the spectral properties of the underlying networks. At a very high assortativity regime, even with a slight decrease in the value of $r$, the Brody parameter smoothly turns one. A further decrease in the values of $r$ leads
to an increase in the value of $L_0$ of $\Delta_3(L)$
statistic for which the spectra follows GOE statistics. As Brody parameter $\beta$ captures the changes in assortativity coefficient at a fine scale \cite{SJ_pre2007a} and $L_0$ at large scale \cite{SJ_epl2009}, which further suggest that when $r$ decreases, randomness increases. With a further decrease in $r$, at around $r=0$, the density distribution start exhibiting peak at zero eigenvalue which becomes more pronounced as $r$ decreases further. Interestingly, most of the studies on network spectra report that the bulk part of the spectra of the networks having Gaussian and scale-free degree distribution follow semi-circular 
and triangular distributions \cite{Farkas_PRE_2001,Aguiar} respectively, but for highly disassortative networks, the spectral density of both the degree distributions can have entirely different behaviour.

Recently, the realm of assortativity has been realized in understanding adaptive synchronization \cite{Gaytan}, which combined with our results of varying amount of randomness for various values of $r$ can be explored further to understand dynamical processes on networks.
Further, Table I indicates that disassortativity is one of the factors contributing to degeneracy at zero. Prevalence of zero degeneracy has been implicated in terms of gene duplication \cite{Gene_duplication}. This, along with the impact of change in topology of a network, brought upon by assortativity, leading to a profound change in the spectral density, provides a direction to explore the evolutionary origin of real world systems \cite{vaccination,migration}.
Lastly, since randomness or random 
connections in a network has already been
emphasized for proper functioning of corresponding systems \cite{rand_realsys}, the profound role
of assortativity parameter revealed through the sophisticated random matrix
technique is not only important for network community attempting to model complex systems, but is interesting for
random matrix communities at the fundamental level as well.

\section{Appendix}
\label{appendix}
Numerical calculations pertaining to assortative mixing, eigenvalues calculations and $\Delta_3(L)$ statistic are done using FORTRAN code written by the Authors. The eigenvalues are calculated by calling LAPACK (Linear Algebra PACKage) subroutines into the FORTRAN code. The calculation of spacings and polynomial fittings are done using MATLAB.

\begin{figure}[t]
\centerline{\includegraphics[width=0.9\columnwidth,height=6cm]{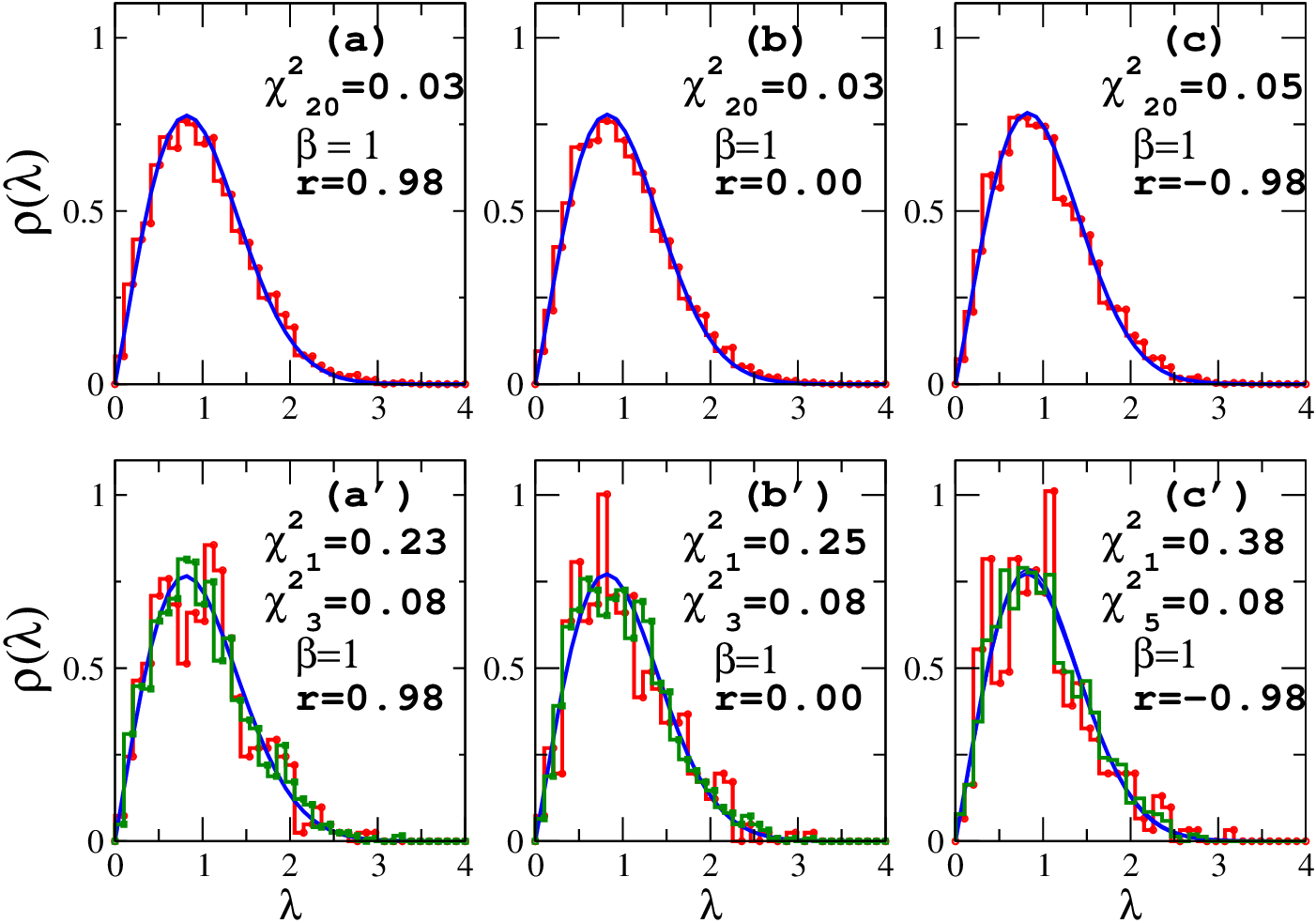}}
\caption{(Color Online) (a), (b) and (c) plot average NNSD for an ensemble of twenty realizations for different values of $r$, whereas ($a\acute{}$), ($b\acute{}$) and ($c\acute{}$) plot the ensemble having different number of the network realizations. The histogram is drawn using 
the data fro the networks, and the solid line is the fitted Brody distribution. For all the graphs $N=1000$ and $\langle k \rangle=10$.}
\label{chi}
\end{figure}

 We present the $\chi^2$ values as a measure of goodness of fit of the model to data, a lower of $\chi^2$ indicating a better fitting. As depicted from Fig.~\ref{chi}, 
the $\chi^2$ values consistently decrease with an increase in the number of network realizations in the ensemble implicating increase in the accuracy reaching to the 
value of $\chi^2$ being less than one lying in the acceptable range \cite{chi}. For assortative networks, as less as three realizations in the individual ensemble are good 
enough to bring $\chi^2$ within the acceptable range, whereas for $r$ taking negative values, the number of realizations in the ensemble increases little bit more (five as depicted in Fig.~\ref{chi}($c\acute{}$)) in order to bring $\chi^2$ within the acceptable range. This happens as for disassortative networks, there is high degeneracy at zero eigenvalue leading to less effective dimension of the unfolded spectra (refer discussions in the Methods and Techniques section) and hence more number of realizations of the networks are required.

\section{Acknowledgements} 
SJ thanks Department of Science and Technology (DST), Govt. of India grant
SR/FTP/PS-067/2011 and Council of Scientific and Industrial
Research (CSIR), Govt. of India grant 25(0205)/12/EMR-II for funding. It is pleasure to acknowledge 
Dr. Kunal Bhattacharya (BITS PILANI) and Igor M. Sokolov (Humboldt-Universit\"{a}t zu Berlin) for
help with the Sokolov's algorithm. AY acknowledges Council of Scientific and Industrial
Research (CSIR), Govt. of India for financial support.

\end{document}